\documentclass[11pt]{article}

\usepackage[margin=1in]{geometry}

\usepackage{graphicx} 
\usepackage{amsmath,amssymb,amsthm,easybmat,mathtools,dsfont}
\usepackage{xcolor}
\usepackage[numbers,sort&compress]{natbib}
\usepackage{hyperref}
\hypersetup{colorlinks=true, linkcolor=blue!80!black, urlcolor=blue!80!black, citecolor=blue!80!black}

\usepackage{comment}
\usepackage{outlines}
\usepackage{subcaption}

\usepackage{authblk}
\theoremstyle{definition}

\newcommand{\ket}[1]{|#1\rangle}

\title{Routing Entanglement in Complex Quantum Networks Using GHZ States}

\author[1]{Xin-An Chen}
\author[2]{Caitao Zhan}
\author[2]{Joaquin Chung}
\author[2]{Jeffrey Larson}

\affil[1]{Department of Electrical and Computer Engineering, University of Illinois at Urbana-Champaign, Urbana, IL, USA}
\affil[2]{Argonne National Laboratory, Lemont, IL, USA}

\date{} 

\begin{document}

\maketitle

\begin{abstract}
    Distributing entanglement to distant parties in a network is a central task in quantum information processing and quantum networking. The sensitivity of entangled states to loss necessitates the use of entanglement routing strategies.
    Recently, a routing strategy using Greenberger-Horne-Zeilinger (GHZ) measurements instead of Bell state measurements (BSM) has been proposed. We further this direction of research by explicitly considering the varying measurement success probabilities of GHZ measurements. Moreover, we extend the analysis beyond square grid networks to complex network models such as Waxman networks and scale-free networks, as well as SURFnet, a real-world network topology in the Netherlands. Taking into account the varying success probabilities, na\"ive application of GHZ routing achieves rates much lower than the conventional BSM routing. Instead, we propose a hybrid GHZ-BSM routing strategy. The hybrid GHZ-BSM routing strategy outperforms BSM routing in square grid networks. In other networks, however, more sophisticated adaptations using global information are required.
\end{abstract}

\section{Introduction}
Entanglement is an essential resource for realizing many quantum information processing tasks, including quantum key distribution \cite{ekert1991quantum}, distributed quantum sensing \cite{zhang2021distributed}, and distributed quantum computing \cite{caleffi2024distributed}. Distributing entanglement to parties separated by significant distance is, therefore, a central task in quantum information science. Because of the lossy nature of optical fibers, however, distributing entanglement through fibers remains challenging, and distribution over long distances requires the use of quantum repeaters \cite{azuma2023quantum} and entanglement routing strategies \cite{abane2025entanglement}. 

Entanglement routing in quantum networks involves finding an optimal strategy to connect intermediate, short-distance entanglements into long-range ones. On a network described by an undirected graph $G=(V,E)$, where $V$ represents the nodes of the network and $E$ represents the physical connections between nodes, conventional end-to-end entanglement routing methods begin by generating Bell states between each pair of connected nodes. Then, paths between two end users are selected, and each node along these paths performs Bell state measurements (BSMs) to connect the Bell states, resulting in end-to-end entangled states between the two end users. In practical scenarios, generation of elementary links suffers from loss, and the BSMs succeed only probabilistically. Pant et al.~\cite{pant2019routing} studied entanglement routing under this setup. The authors introduced and evaluated the performance of a greedy method to select the paths. 

Since each BSM succeeds with probability less than one, as the distance between two end users (and therefore the number of hops between them) increases, the rates of entanglement routing protocols based on BSM will decrease, typically in an exponential fashion. In Ref.~\cite{patil2022entanglement}, a novel protocol based on Greenberger--Horne--Zeilinger (GHZ) measurements was proposed. The authors identified a close relationship between the performance of their proposed protocol and the phenomenon of percolation \cite{grimmett1999percolation}, which allowed them to conclude that as long as Bell state generation and the GHZ measurements succeed with probabilities above a critical threshold, the rate of the protocol will be distance-independent.
One potential drawback of this GHZ routing protocol is the assumption that $k$-qubit GHZ measurements succeed with the same probability for all $k$, which can be unjustified in practice. Additionally, the authors focused primarily on square grid networks, without touching on more complex network models. 

In our work, we explicitly consider the varying success probabilities of GHZ measurements. First, we model the success probability of $k$-GHZ measurements as an exponentially decreasing function $q^{k-1}$. Alternatively, we assume we constrain ourselves to only two-qubit (i.e., BSM) and three-qubit GHZ measurements. As we will demonstrate in Section \ref{sect:results}, GHZ routing performs poorly if either assumption is adopted. To this end, we propose a hybrid GHZ-BSM routing strategy. Intuitively, this hybrid strategy simulates GHZ measurements by local preparation of a GHZ state and BSM. 
We numerically simulate the end-to-end rates achievable with these strategies in three network models, namely, square grid, Waxman \cite{waxman1988routing}, and scale-free networks \cite{barabasi1999emergence,yook2002modeling}, as well as the topology of SURFnet, a real-world network for research and education (R\&E) in the Netherlands \cite{topology-zoo}. We explicitly demonstrate that in square-grid networks, hybrid GHZ-BSM routing exhibits distance-independence and achieves higher rates than BSM routing, as long as the measurement success probability is above a critical threshold. 
In other network topologies that we considered, a na\"ive implementation of the GHZ-based protocols fails to yield obvious advantages. %
This result suggests that implementing GHZ-based protocols in realistic networks requires more sophistication and coordination among the nodes. In particular, we propose using the idea of network segmentation to divide the network into smaller regions (analogous to OSPF areas in classical routing), with each region of the network performing entanglement routing in a parallel fashion. This strategy can significantly increase the rates achievable with the two GHZ-based protocols.
Moreover, the use of regions can unlock hierarchical routing strategies that are more scalable than state-of-the-art approaches.

\section{Preliminaries}
This section introduces the necessary background on the network topologies used in our evaluation, as well as the state of the art in entanglement routing using GHZ measurements.

\subsection{Network Models}
In this work we will frequently distinguish between physical network topologies and virtual network topologies. A physical network topology, described by a graph $G_\mathrm{phys}=(V,E_\mathrm{phys})$, consists of network nodes and physical connections between them, such as optical fibers or communication channels. A virtual topology $G_\mathrm{vir}=(V,E_\mathrm{vir})$, on the other hand, consists of the established entangled links between two adjacent nodes. We will compare the performance of entanglement routing protocols on three widely adopted network models, which we describe below. For each of these models, we assume that the physical network is deployed over a square region $\Omega=[0,R]\times[0,R]$.

\textit{Square Grid Networks.} In a square grid network, $N\times N$ nodes are placed on the vertices of a square grid. Neighboring nodes are separated by a distance $R/(N-1)$.

\textit{Waxman Networks.} In a Waxman network \cite{waxman1988routing}, $N$ nodes are sampled uniformly at random on $\Omega$. Each pair of nodes is connected with a probability $p(x_i,x_j)=\beta\exp(-\|x_i-x_j\|/\alpha L)$, where $L=\sqrt{2}R$. For the fiber optical network in the United States, the parameters are estimated to satisfy $\beta=1$ and $\alpha L=226$ km \cite{lakhina2003geographic,durairajan2015intertubes}.

\textit{Scale-Free Networks.} Another interesting network model we consider is scale-free networks \cite{barabasi1999emergence,yook2002modeling}. These networks are built dynamically according to the principle of preferential attachment. We begin with an initial network of $m_0$ nodes placed on $\Omega$. At each step, a new node, whose position $x_i$ is chosen uniformly at random, is added to the network and connects to $m$ existing nodes. The probability that the new node is connected to node $j$ located at $x_j$ is $p(x_i,x_j) \propto k_j^\mu/||x_i-x_j\|^\nu$, where $k_j$ is the current degree of node $j$ and $\mu,\nu$ parameterize the strength of preferential attachment and distance effects, respectively. These networks are called scale-free because the degree of a node  in the network follows a simple power law distribution $p(k)\propto k^{-\gamma}$ and is independent of any notion of scale in the network. For simplicity, we  choose $\mu=\nu=1$ in the present work.

\begin{figure*}
    \centering
    \subfloat[Physical Waxman network]{\includegraphics[width=0.33\textwidth]{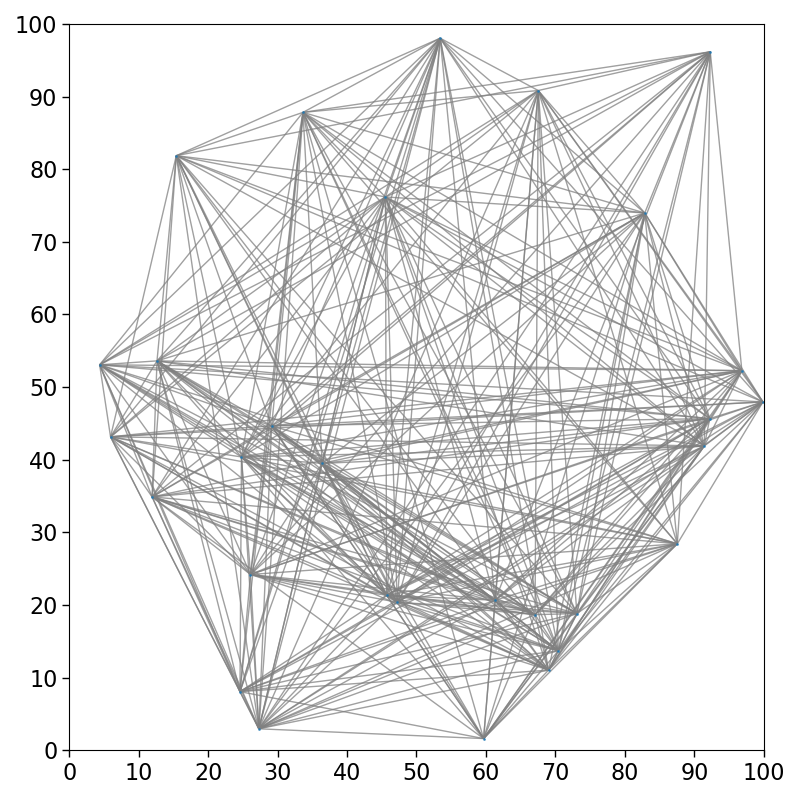}}
    \subfloat[Physical scale-free network]{\includegraphics[width=0.33\textwidth]{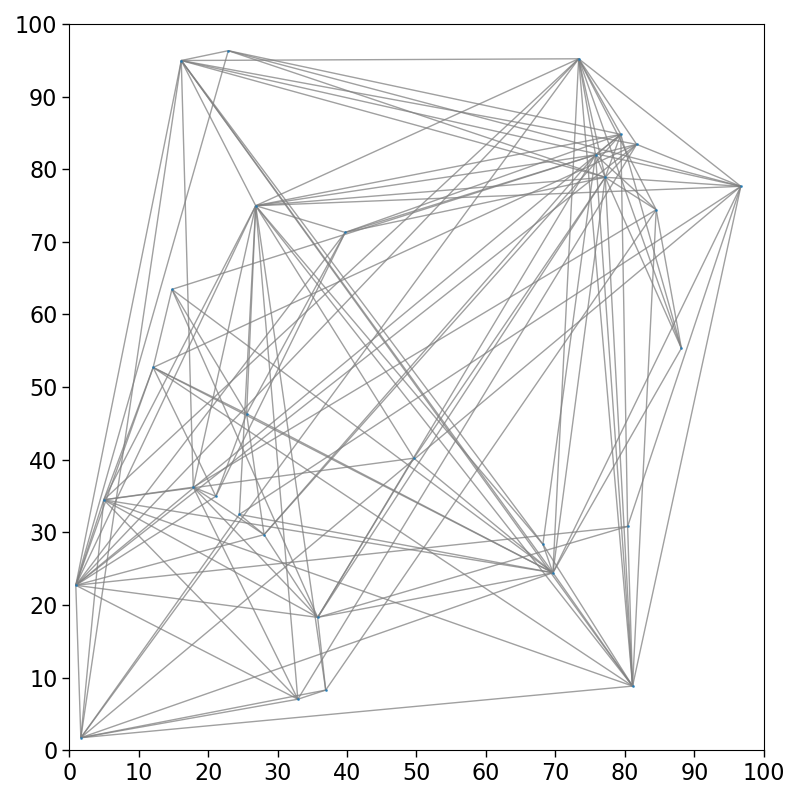}}
    \subfloat[SURFnet topology]{\includegraphics[width=0.33\textwidth]{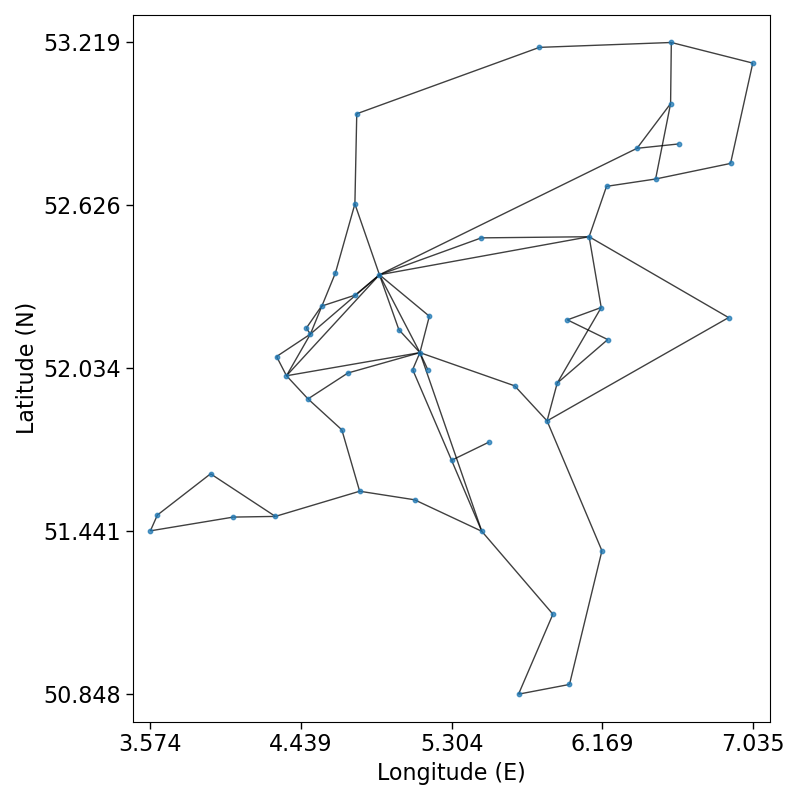}\label{fig:surfnet}}
    \caption{(a) Example of a physical Waxman network with 30 nodes on a 100 km $\times$ 100 km square. We chose $\alpha=1.598$ so that $\alpha L=226$ km, and $\beta=1$. (b) Example of a scale-free network with 30 nodes on a 100 km $\times$ 100 km square. We chose $m=5$, $\mu=\nu=1$, and the initial network is a 6-node complete graph. (c) SURFnet topology}
    \label{fig:physical_networks}
\end{figure*}

We will henceforth fix $R=100$ km, so that the underlying region is always a $[0,100\text{ km}]\times[0,100\text{ km}]$ square. A visualization of the Waxman model and the scale-free model is shown in Fig.~\ref{fig:physical_networks}. We can readily identify an important distinction between these two models. In scale-free networks  a few highly connected ``hubs" exist, whereas the Waxman network is much more homogeneous. 

In addition to the three network models presented above, we study the performance of routing protocols on the topology of SURFnet, a real-world network used for research and educational purposes in the Netherlands. Its topological information can be obtained from the Internet Topology Zoo \cite{topology-zoo}. Figure \ref{fig:surfnet} provides a visualization of the SURFnet topology.

\begin{figure*}
    \centering
    \subfloat[Virtual Waxman network]{\includegraphics[width=0.36\textwidth]{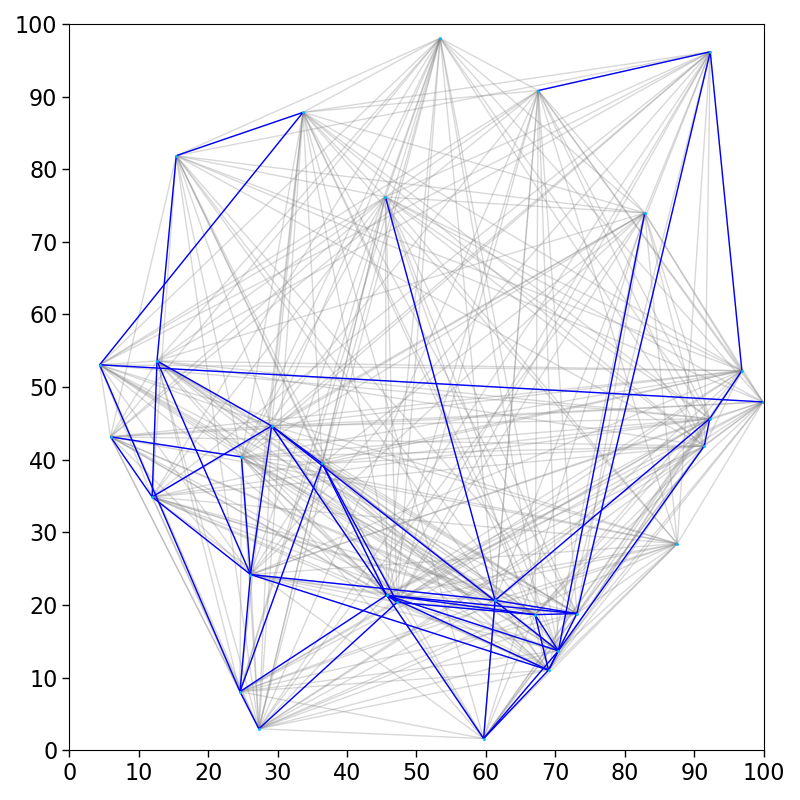}}
    \subfloat[Virtual scale-free network]{\includegraphics[width=0.36\textwidth]{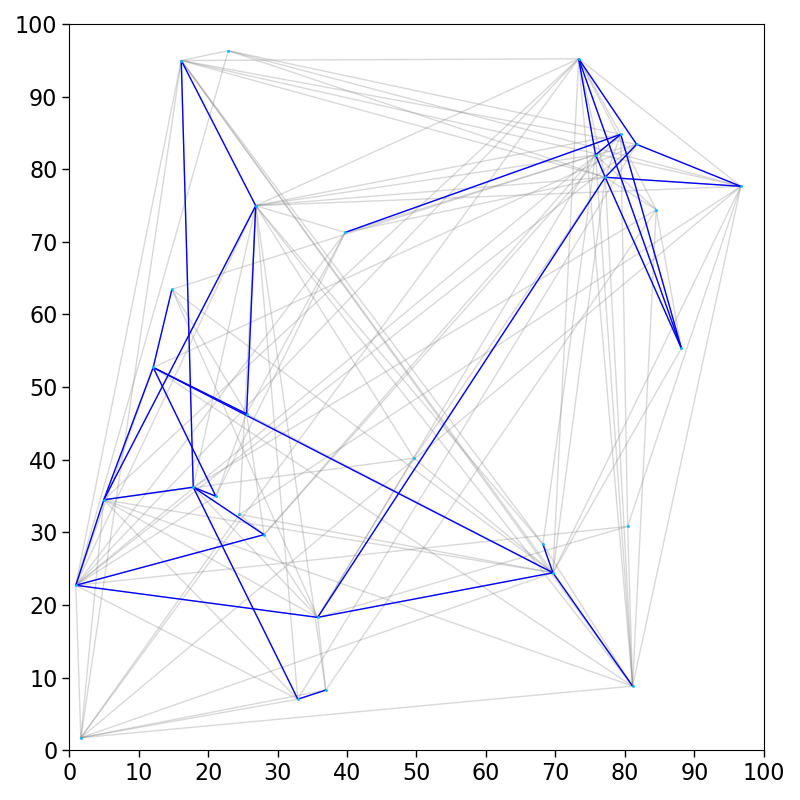}}
    \caption{(a) Example of a virtual Waxman network. (b)  Example of a virtual scale-free network. The gray lines represent physical links, while the blue lines represent the successfully established entanglement links between two adjacent nodes. For both, we have assumed that the fiber-optic loss is given by $\gamma=0.2$ km$^{-1}$.}
    \label{fig:virtual_networks}
\end{figure*}

We assume that the physical connections between nodes are optical fibers, such that the loss is given by $\eta=10^{-\gamma d/10}$, where $\gamma$ is typically 0.2 km$^{-1}$ and $d$ is the distance of the transmission.
Figure ~\ref{fig:virtual_networks} shows a virtual topology overlaid on top of the physical topology for both Waxman and scale-free network models.


\subsection{Related Work}\label{sect:relevant-works}
A plethora of literature exists on the topic of entanglement routing, bringing together ideas from quantum information theory, statistical physics, and network science. In this subsection we provide a short overview of a few relevant works. 

A fundamental upper bound on end-to-end quantum capacities in a quantum network was first established by Pirandola, providing an ultimate performance limit for entanglement routing protocols \cite{pirandola2019end}. Others \cite{acin2007entanglement,brito2020statistical,zhuang2021quantum} have observed phase transitions in the connectivity of quantum networks. Below a critical node density, a quantum network consists of disconnected ``islands,'' and the end-to-end capacities between most nodes are zero. Above the critical density, the network becomes sufficiently connected, and the end-to-end capacities increase with more nodes. On the more practical side, in Ref. \cite{harney2025practical}, the authors proposed algorithms for optimizing single-path and multipath entanglement routing in practical scenarios with an emphasis on resource efficiency, and they numerically demonstrated that a multipath routing protocol performs better on Waxman and scale-free networks.

The aforementioned studies assume that the entanglement swapping operation succeeds deterministically. In practice, however, this assumption often does not hold. Along these lines, Pant et al.~\cite{pant2019routing} analyzed the performance of entanglement routing with probabilistic entanglement swapping and found that in this setting a multipath routing protocol on a square grid performs much better than if the users are  connected only by a one-dimensional repeater chain. Specifically, they considered the following scenario. In the first step, generation of Bell state links between each neighboring nodes is attempted, which succeeds with probability $p$ for all edges. Next, the state of each link (i.e., whether the attempt was successful) is sent to a central server that decides the paths to route the entanglement to the two end users. These selected paths are revealed to the relevant nodes, which will perform (probabilistic) BSM accordingly. The authors proposed a greedy method for selecting the paths. The central server first selects the shortest path between the two end nodes. This path is then eliminated from the graph, and the algorithm selects the shortest path in the pruned graph. This procedure is iterated until  no more paths exist between the two end nodes. In the following, we will refer to this protocol as \textbf{BSM routing}.

Note that BSM routing requires each node to report to a central server whether entanglement generation has succeeded; in turn, the central server needs to inform all nodes how to perform the routing. This potentially introduces more latency into the routing protocol due to classical coordination. In Ref. \cite{patil2022entanglement}, 
an alternative protocol based on GHZ measurements is proposed. Instead of asking a central server to find the shortest paths greedily, the nodes blindly perform GHZ measurements based on their local information alone. In more detail, after Bell state generation attempts, all helper (i.e., non-user) nodes that have $k\geq1$ connections with neighbors perform a $k$-qubit GHZ measurement on the $k$ qubits that they hold. If $k=1$, the $k$-qubit GHZ measurement reduces to an $X$ measurement, which disentangles the node from the rest of the network. If $k=2$, the measurement reduces to a BSM. The $k$-GHZ measurements are assumed to be successful with probability $q$ for all $k\geq2$ for all nodes. 
Following Ref. \cite{bartolucci2023fusion}, failure of the GHZ measurements is modeled by a single qubit stabilizer measurement, which can be modified into an $X$ measurement by a simple change of basis. The performance of this protocol is directly connected to the phenomenon of percolation \cite{grimmett1999percolation} in statistical physics; and the authors found that when $p$ and $q$ are above a certain percolation threshold, then almost the entire network is connected in the limit that the number of nodes $n\to\infty$, and the protocol almost always yields exactly one Bell state between the user nodes in each cycle, regardless of how far apart they are. On the other hand, the authors showed that the rate achievable with BSM alone is at most $O(q^d)$ on square grid networks, where $d$ is the Manhattan distance between the nodes. In a later work, numerical evidence was provided to show that the rate no longer remains distance-independent when the initial Bell state links are imperfect \cite{kaur2023distribution}. Nevertheless, it remains open how the performance of the GHZ measurement protocol and the BSM protocol compare.

\section{GHZ routing with realistic assumptions and design of hybrid GHZ-BSM routing}
The analysis in Ref. \cite{patil2022entanglement} assumes $k$-GHZ measurements succeed with the same probability for arbitrarily large $k$. We will henceforth call this setting \textbf{uniform-success GHZ routing}. In matter qubit systems, $k$-GHZ measurements simply consists of a $k$-qubit rotation from the GHZ basis to the computational basis followed by a computational basis measurement. Therefore, all $k$-qubit measurements succeed with unit probability. In photonic systems, however, BSM succeeds probabilistically \cite{calsamiglia2001maximum,grice2011arbitrarily}, and we expect $k$-GHZ measurements have decreasing success probability for increasing $k$. As such, we will explicitly take into account the varying success probabilities of GHZ measurements. Our first approach is to assume that the success probability of $k$-GHZ measurements decrease exponentially as $q^{k-1}$. GHZ routing performed under this setting will be called \textbf{exponential-decay GHZ routing}. In our second approach, we constrain ourselves to BSM and 3-GHZ measurements only, both of which are assumed to succeed with probability $q$. We will call this setting \textbf{(2,3)-GHZ routing}. 

\begin{figure}
    \centering
    \includegraphics[width=0.5\linewidth]{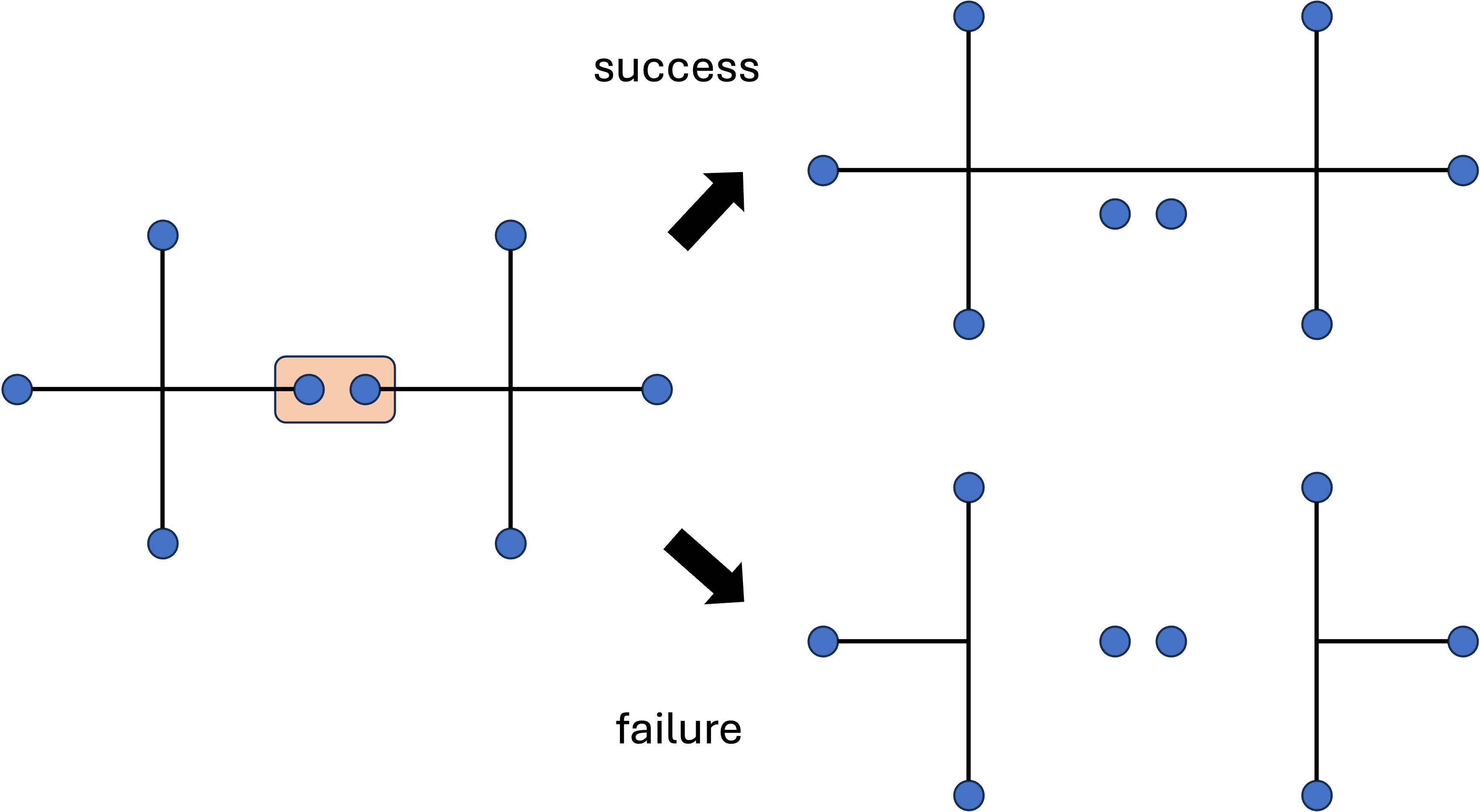}
    \caption{Illustration of the output states after a Bell state measurement}
    \label{fig:bsm_illustration}
\end{figure}

We furthermore propose a modified protocol based on preparation of GHZ states and BSM. In more detail, each cycle of our protocol is divided into three steps:
\begin{enumerate}
    \item \textbf{Bell state generation.} Each end-user node generates a Bell state with each of its neighbors. Transmission from node $i$ to node $j$ succeeds with probability $\eta_{ij}=10^{-\gamma d_{ij}/10}$, where $d_{ij}$ is the distance between $i$ and $j$ and $\gamma=0.2$ km$^{-1}$, corresponding to 0.2 dB/km loss in current optical fibers. Every node that is not an end user acts as a helper. 
    \item \textbf{GHZ state generation.} Each helper node furthermore generates a $k$-GHZ state $\ket{\mathrm{GHZ}_k}=\frac{1}{\sqrt{2}}(\ket{0}^{\otimes k}+\ket{1}^{\otimes k})$, where $k$ is the degree of the node in the virtual network. The GHZ states are stored locally at the helper nodes.
    \item \textbf{Bell state measurement.} Each helper performs a BSM, succeeding with probability $q$. As before, a measurement failure results in two GHZ states with one less qubit (see Fig.~\ref{fig:bsm_illustration}).
    \item \textbf{Classical communication.} The measurement results are sent to the two user nodes, which will apply a Pauli unitary accordingly.
\end{enumerate}

\begin{figure}
    \centering
    \hfill
    \subfloat[]{\includegraphics[width=0.4\linewidth]{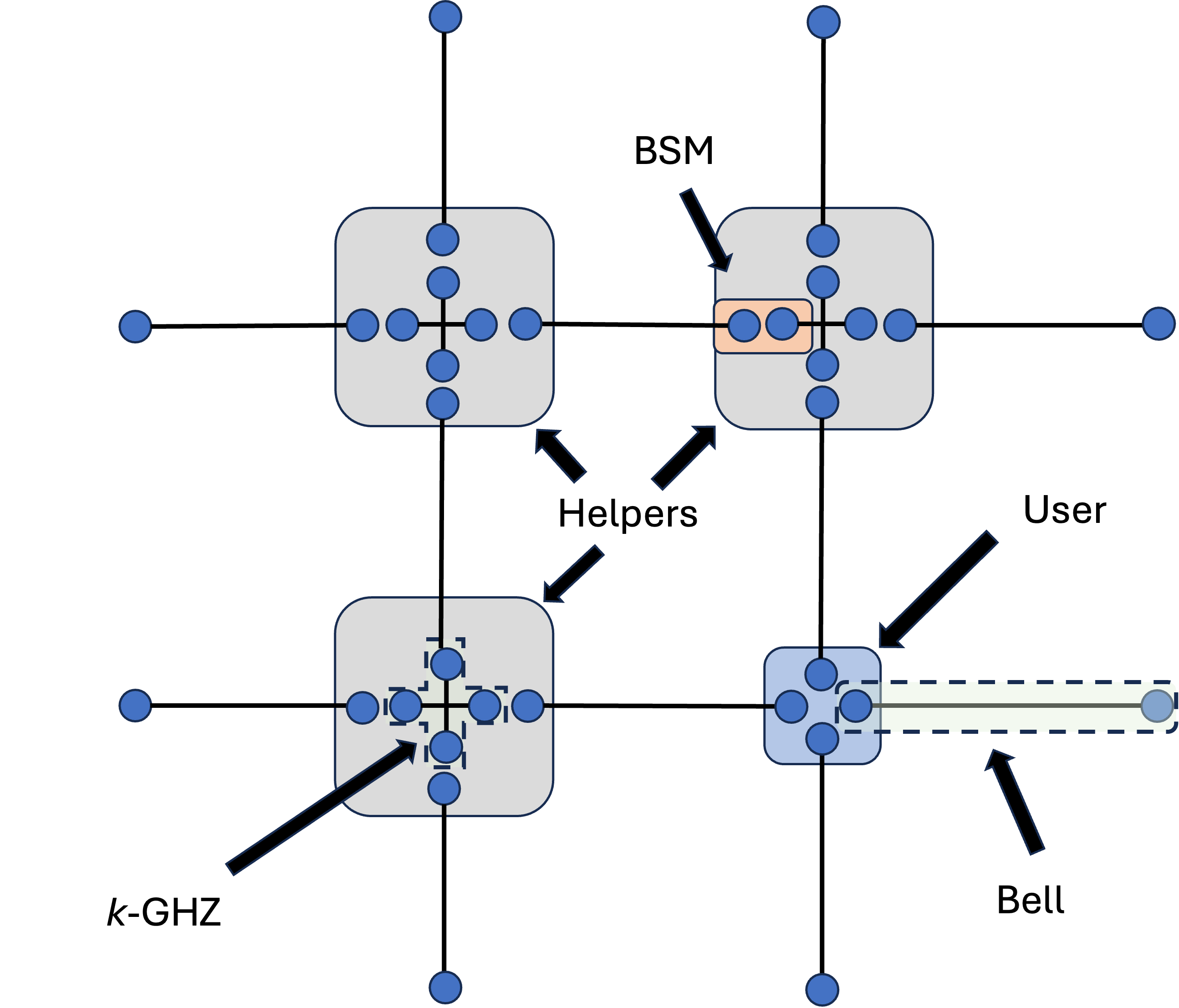}} 
    \hfill
    \subfloat[]{\includegraphics[width=0.45\linewidth]{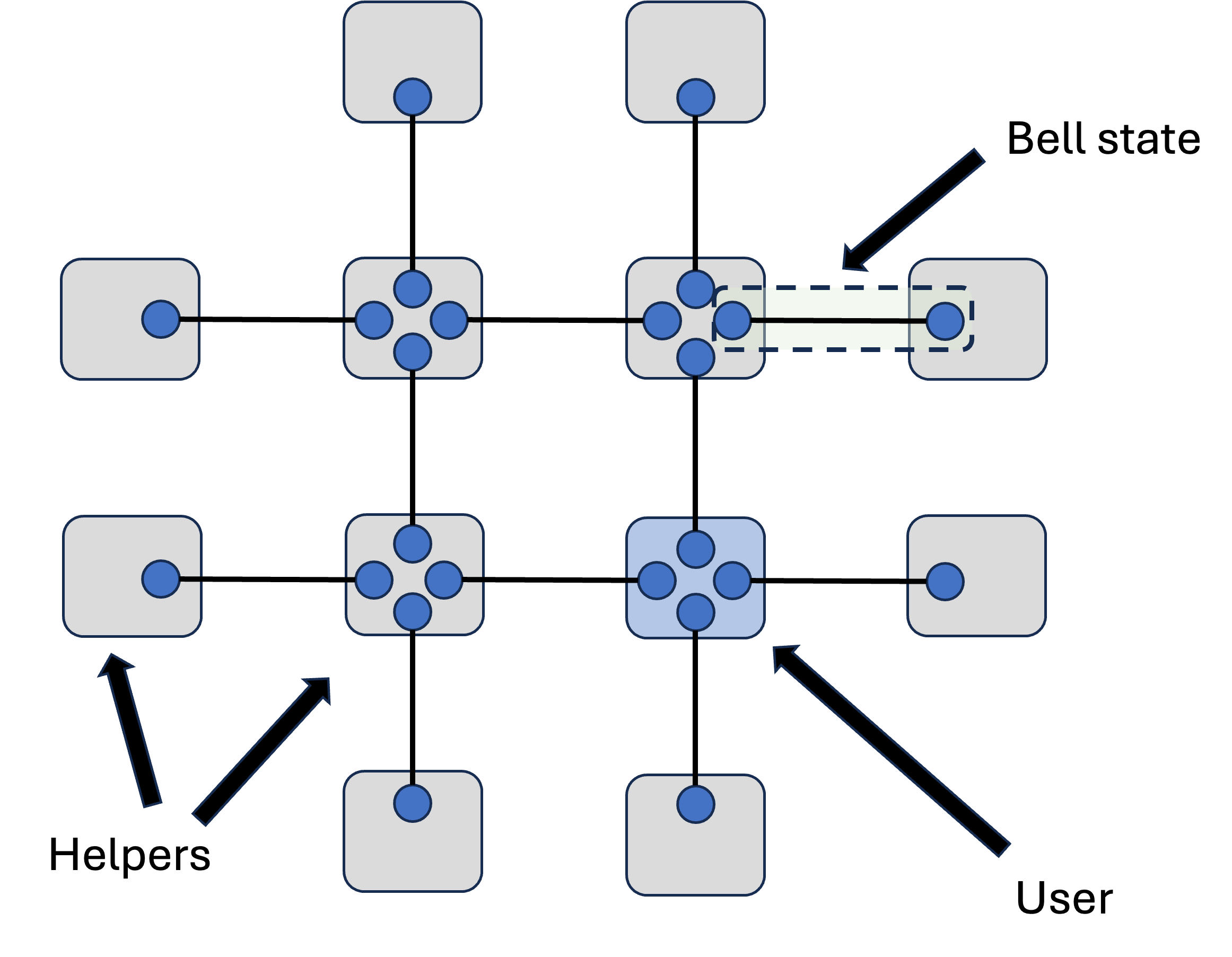}}
    \hfill
    \caption{(a) Illustration of our proposed GHZ state protocol on a square grid. The blue box represents one of the two end users, the gray boxes represent helper nodes, the green shaded region represents the prepared states, and the orange boxes represent a Bell state measurement on the two received photons. (b)  Illustration of the original GHZ measurement protocol \cite{patil2022entanglement,kaur2023distribution}. The blue box represents one of the two users, and the gray boxes represent helpers that perform $l$-GHZ measurements, where $l$ is the number of qubits that is received by the user.}
    \label{fig:protocol_illustration}
\end{figure}

The transmission between nodes $10^{-\gamma d/10}$ may be too low for large $d$. In this case, we may try to improve transmission probability by increasing the number of attempts. Our protocol, which we will call \textbf{hybrid GHZ-BSM routing}, is illustrated in Fig.~\ref{fig:protocol_illustration}, together with a graphical comparison with the the GHZ routing strategy from Ref.~\cite{patil2022entanglement}. 

\section{Methods}
The performance of the protocols is measured by their end-to-end rates, defined as the expected number of end-to-end Bell states obtained per cycle. More specifically, we will consider two  figures of merit, the average rate over all node pairs $\langle R\rangle$, given by
\begin{align}
    \langle R\rangle=\binom{|V|}{2}^{-1}\sum_{u,v\in V} R(u,v),
\end{align}
where $R(u,v)$ is the rate between nodes $u$ and $v$, and the dependence of $R(u,v)$ on the distance $d(u,v)$ between $u$ and $v$. In both cases, the rates were obtained by using the Monte Carlo method. 

For Waxman and scale-free networks, we obtain $\langle R\rangle$ by sampling 10 random physical topologies, from which we choose 20 node pairs and compute the rate between these pairs by taking 500 random samples of virtual topologies. For square grids, since there is no randomness in the physical topology, we simply choose 100 node pairs and again compute the rate between these pairs by sampling 500 virtual topologies on the square grid. The obtained average rate will be plotted for varying numbers of nodes. 
For the SURFnet topology, the average rate is obtained by averaging over all node pairs. Since the number of nodes is  fixed, we plot the rate for varying measurement success probability $q$.

To evaluate the distance dependence, we randomly choose 100 node pairs and compute the rate between these pairs by sampling 1,000 virtual topologies. The rates are then plotted against the distance between the node pairs.

\section{Evaluation Results}\label{sect:results}

\begin{figure*}
    \centering
    \subfloat[]{\includegraphics[width=0.25\linewidth]{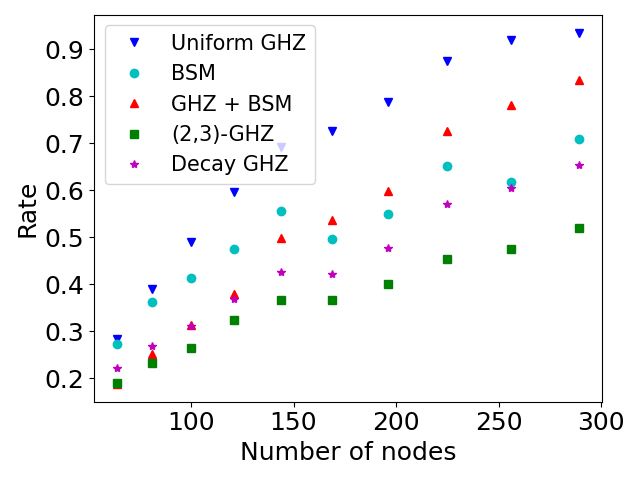}\label{fig:square-avg-rate}}
    \subfloat[]{\includegraphics[width=0.25\linewidth]{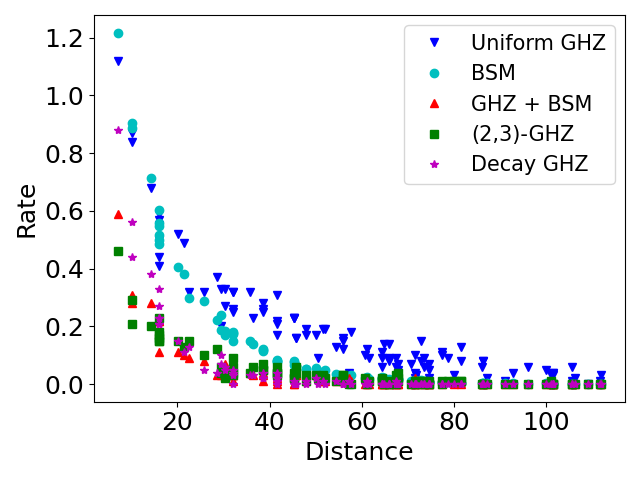}\label{fig:square-rate-distance-0.7}}
    \subfloat[]{\includegraphics[width=0.25\linewidth]{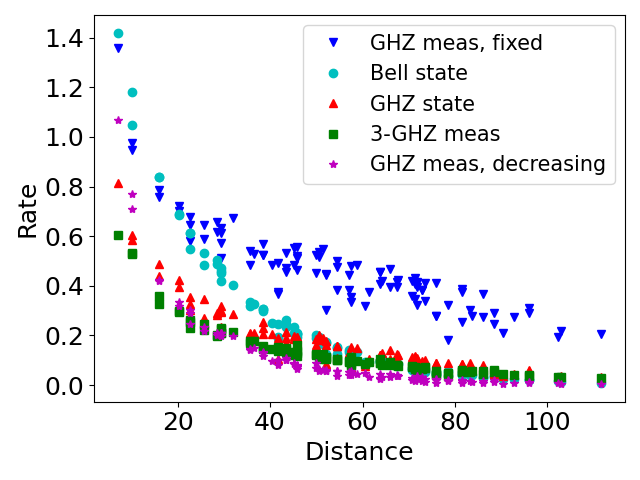}}
    \subfloat[]{\includegraphics[width=0.25\linewidth]{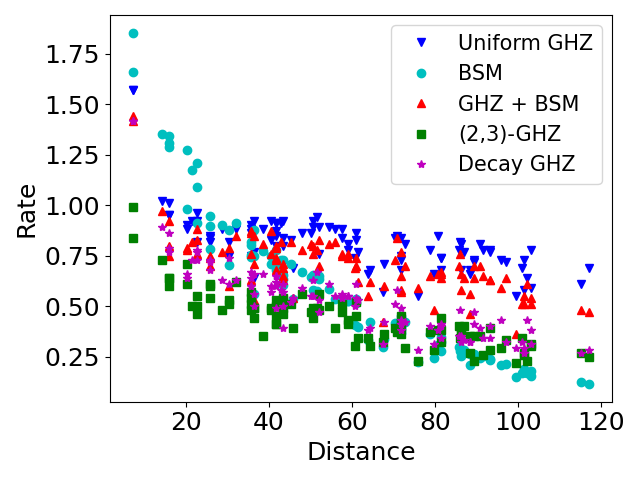}\label{fig:square-rate-distance-0.9}}
    \caption{Performance evaluation for square grid networks. (a) Average rate $\langle R\rangle$ for an $N\times N$ grid for $N$ from 8 to 17. (b--d) Rate vs. distance (in km) for $N=15$ and $q=0.7,0.8,0.9$, respectively.}
    \label{fig:square-grid-results}
\end{figure*}
\begin{figure*}
    \centering
    \subfloat[]{\includegraphics[width=0.25\linewidth]{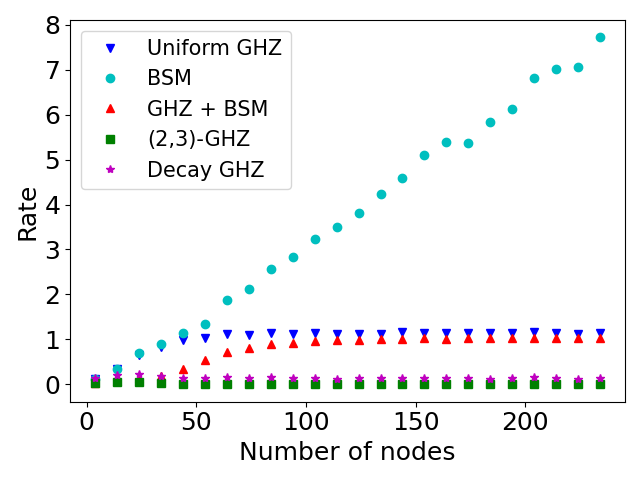}}
    \subfloat[]{\includegraphics[width=0.25\linewidth]{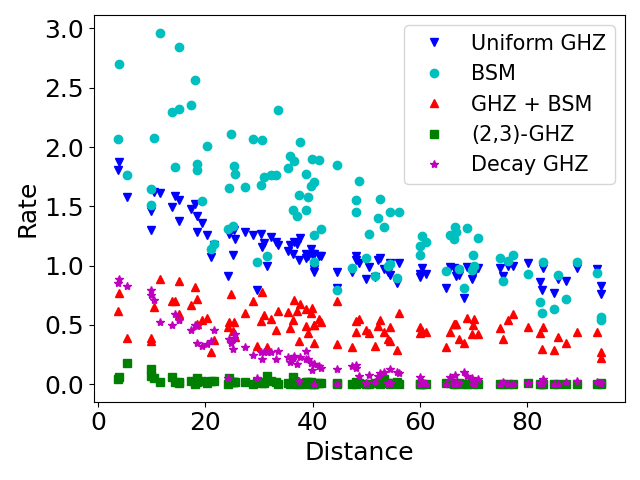}}
    \subfloat[]{\includegraphics[width=0.25\linewidth]{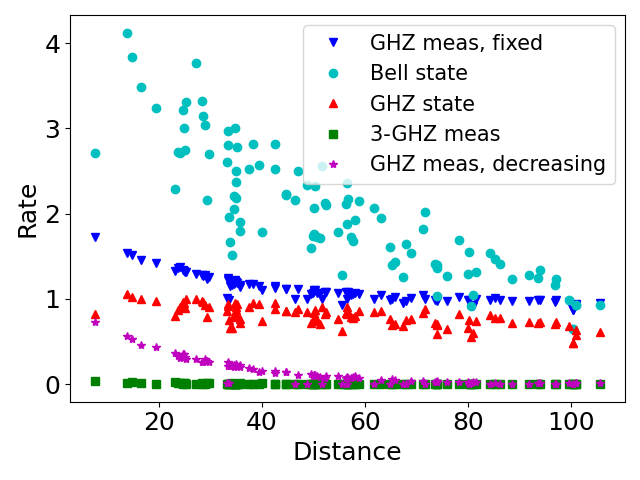}}
    \subfloat[]{\includegraphics[width=0.25\linewidth]{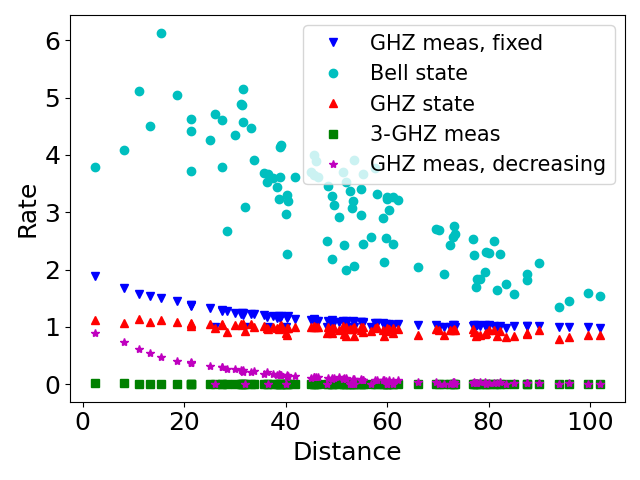}}
    \caption{Performance evaluation for Waxman networks with $q=0.5$. (a) Average rate $\langle R\rangle$ for varying number of nodes. (b--d) Rate vs. distance (in km) for $n=30,50,100$, respectively.}
    \label{fig:waxman-results}
\end{figure*}
\begin{figure*}[]
    \centering
    \subfloat[]{\includegraphics[width=0.25\linewidth]{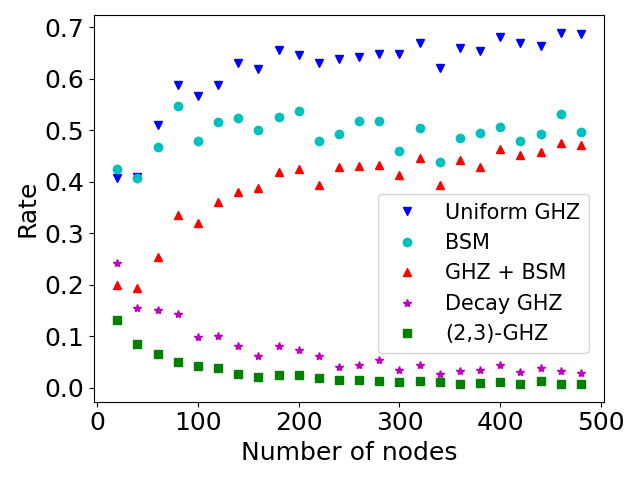}}
    \subfloat[]{\includegraphics[width=0.25\linewidth]{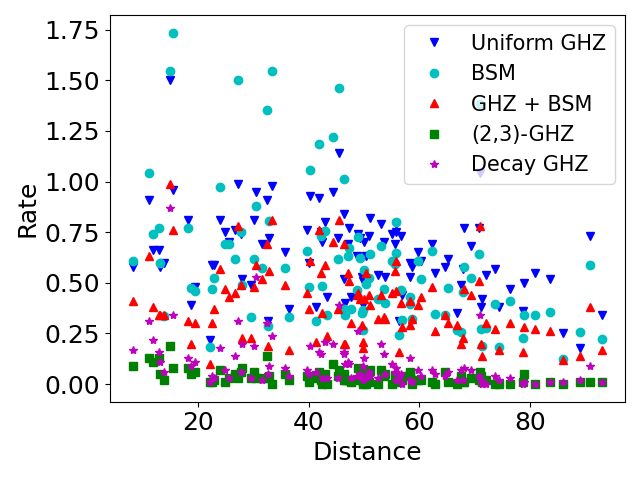}}
    \subfloat[]{\includegraphics[width=0.25\linewidth]{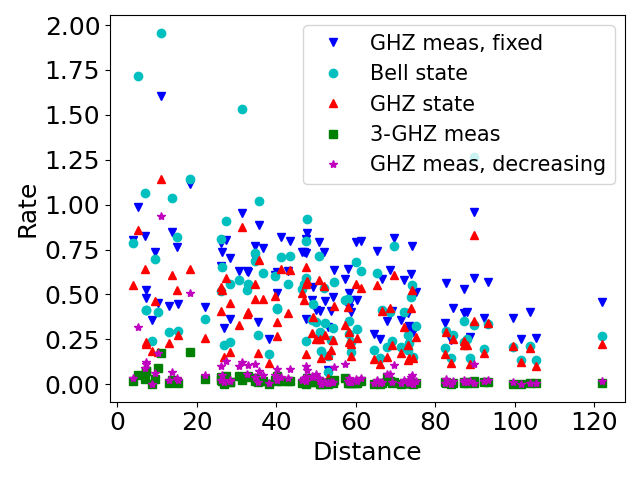}}
    \subfloat[]{\includegraphics[width=0.25\linewidth]{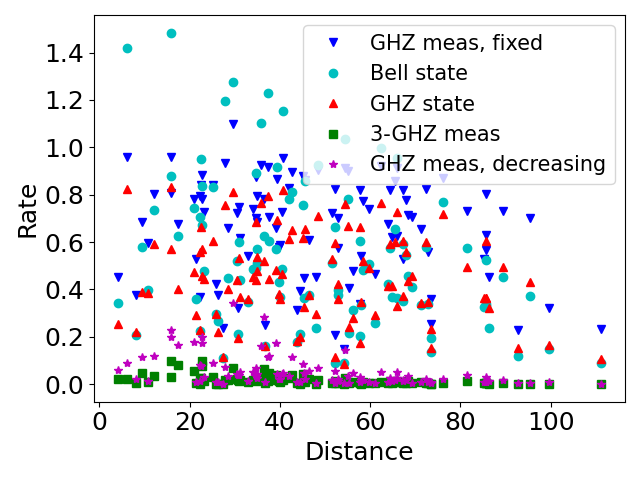}}
    \caption{Performance evaluation for scale-free networks with $q=0.75$. (a) Average rate $\langle R\rangle$ for varying number of nodes. (b--d) Rate vs. distance (in km) for $n=100,150,250$, respectively.}
    \label{fig:sf-results}
\end{figure*}
\begin{figure*}[]
    \centering
    \subfloat[]{\includegraphics[width=0.25\linewidth]{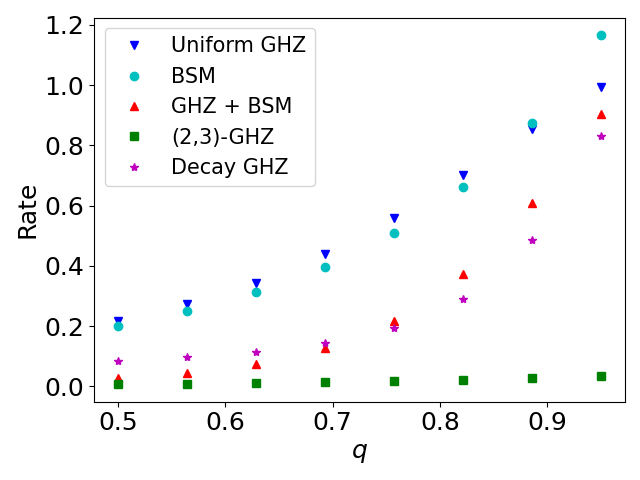}}
    \subfloat[]{\includegraphics[width=0.25\linewidth]{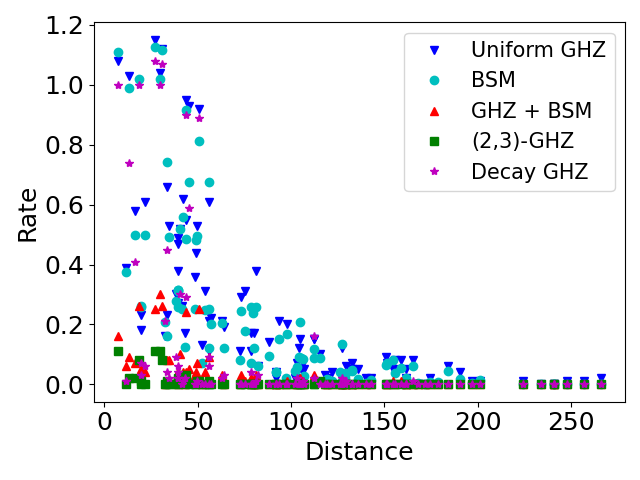}}
    \subfloat[]{\includegraphics[width=0.25\linewidth]{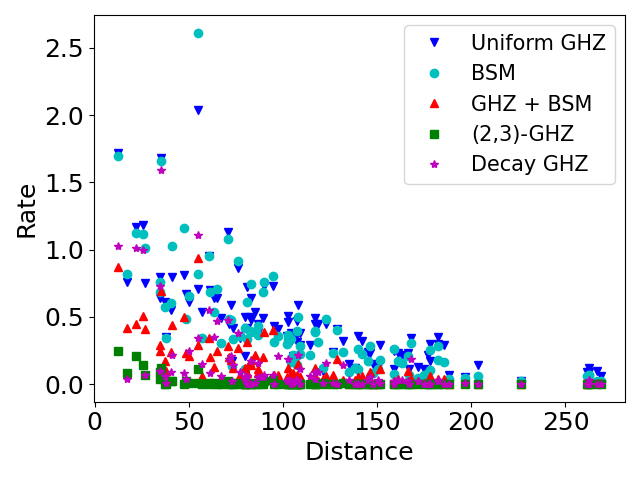}}
    \subfloat[]{\includegraphics[width=0.25\linewidth]{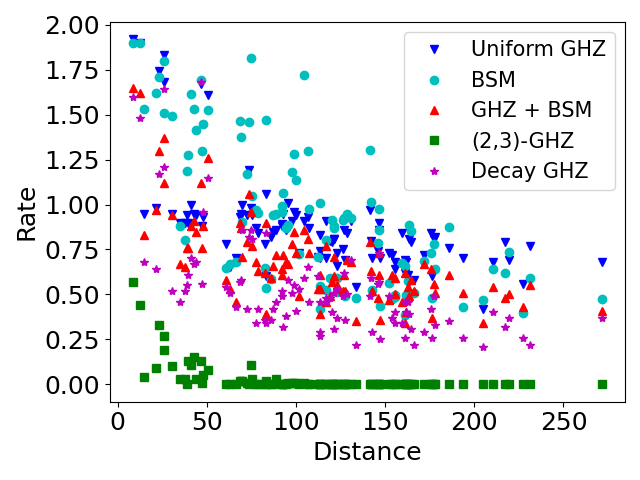}\label{fig:surfnet-rate-q-0.9}}
    \caption{Performance evaluation on the SURFnet topology. (a) Average rate $\langle R\rangle$ for varying measurement success probabilities $q$. (b--d) Rate vs. distance (in km) for $q=0.5,0.7,0.9$, respectively.}
    \label{fig:surfnet-results}
\end{figure*}

\subsection{Square Grid}

In Ref.~\cite{patil2022entanglement}, the performance of GHZ routing was related to percolation. The network experiences a phase transition between the disconnected phase and the connected phase. If the transmission and measurement success probabilities are not sufficiently high, then the network consists of disconnected ``islands,'' and the probability that two nodes are connected decays exponentially with the distance between them. With sufficiently high transmission and measurement success probabilities, the network enters the connected phase, and the probability that two nodes are connected by a path remains constant with distance. This is the key ingredient in concluding that the rate of uniform-success GHZ routing is distance-independent \cite{patil2022entanglement}. While the Bell state protocol performs similar to or better than the GHZ-based protocols for short distances, the GHZ-based protocols achieve significantly higher rates over large distances.

Reference \cite{patil2022entanglement} provided numerical evidence for the distance independence. In this work we analyze the performance of our proposed GHZ state protocol, while complementing the analysis in Ref.~\cite{patil2022entanglement} by computing the average end-to-end rate and directly observing the rate versus distance behavior of their GHZ measurement protocol. The results are shown in Fig.~\ref{fig:square-grid-results}. In Fig.~\ref{fig:square-rate-distance-0.7}, with the transmission probability given by $\eta=10^{-\frac{100}{10(N-1)}\gamma}$ and measurement success probability $q=0.7$, the network is still in the disconnected phase, and therefore the rates achievable with GHZ routing strategies still decay significantly with distance. As we increase the measurement success probability, however, the network becomes connected, and the GHZ routing strategies achieve distance-independent rates (Fig. \ref{fig:square-rate-distance-0.9}).

Fig. \ref{fig:square-avg-rate} shows the average rates for varying number of nodes in the network. Uniform-success GHZ routing achieves the highest average rate. Hybrid GHZ-BSM routing achieves lower rates than BSM routing does when the number of nodes is small, but performs better than BSM routing when there are sufficiently many nodes in the network. This is expected since the number of hops on a path increases with the number of nodes, resulting in lower rates for BSM routing. While exponential-decay GHZ routing and (2,3)-GHZ routing exhibit distance independence, the average rates they achieve is lower than those achieved by BSM routing.

\subsection{Waxman Networks}
The results for Waxman networks are shown in Fig.~\ref{fig:waxman-results}. In Waxman networks, uniform-success GHZ routing and hybrid GHZ-BSM routing perform better than exponential-decay GHZ routing and (2,3)-GHZ routing, and all GHZ-based strategies all perform significantly worse than the conventional Bell state protocol. These results illustrate a crucial shortcoming of GHZ-based routing: even if multiple paths exist between two nodes, GHZ-based routing make use of these paths in a ``coherent'' fashion, yielding only one pair of end-to-end entangled state at the end in most cases. In contrast, BSM routing adequately makes use of the path redundancy. This is especially significant in Waxman networks for the following reasons. First, the average node degree scales as $O(n)$ with the number of nodes $n$ in Waxman networks \cite{roughan2019estimating}. Therefore, there are $O(n)$ potential paths that can be used in BSM routing. Moreover, these paths tend to be only a few hops, and performing BSM on these paths yields a rate that does not significantly worsen as $n$ increases. The second point can be justified by the following heuristic argument. A crude estimate for the average path length (i.e., the number of steps along the shortest path) between two nodes is $\langle l\rangle\approx\frac{\ln n}{\ln \langle k \rangle}$, where $\langle k \rangle$ is the average degree \cite{smith2007average}. In a Waxman network, the average degree scales linearly, so that $\langle k \rangle \sim \kappa n$. This implies that $\langle l \rangle = O(1)$ in the limit $n\to\infty$. 

Of course, not all hope is lost for the GHZ-based protocols. An important advantage for the GHZ-based protocols is that they  use only local information. BSM routing, in contrast, requires information for the global virtual topology. An interesting follow-up problem is then to design a GHZ-based routing strategy that takes advantage of global information. In the following, we will see that the Waxman networks undergoes a phase transition similar to the square grid network. As the number of nodes increases, the network becomes sufficiently connected; and there is, on average, at least one path between any two nodes. A na\"ive strategy is then to perform network segmentation, in other words, dividing the network into individual segments of critical size, with each segment working in parallel. This will enhance the rate of the GHZ-based routing strategies, such that it also scales linearly with the network size. We leave the detailed analysis to future work.

\subsubsection{Phase transition in Waxman networks.}\label{sect:waxman-phase-transition}
In a Waxman network, the probability that two nodes are connected is 
\begin{align}
    p(x_1,x_2) &= p_c(x_1,x_2)\eta(x_1,x_2) \\
    &= e^{-\|x_1-x_2\|/\alpha L}\left[1-\left(1-10^{-\gamma\|x_1-x_2\|/10}\right)^m\right].
\end{align}
Given $p(x_1,x_2)$, the expected number of paths between two nodes chosen at random is given by \cite{van2001paths}:
\begin{align}
    E_\mathrm{paths} = &\sum_{j=1}^{N-1}\frac{(N-2)!}{(N-j-1)!}\times\\
    &\quad\mathbb{E}_{x_1,x_2,\cdots,x_j}[p(x_1,x_2)p(x_2,x_3)\cdots p(x_{j-1},x_j)],
\end{align}
where 
\begin{align}
    &\mathbb{E}_{x_1,x_2,\cdots,x_j}[p(x_1,x_2)p(x_2,x_3)\cdots p(x_{j-1},x_j)] \notag\\
    &= \int\frac{dx_1}{V}\cdots\int\frac{dx_j}{V} p(x_1,x_2)p(x_2,x_3)\cdots p(x_{j-1},x_j).
\end{align}
Ref. \cite{van2001paths} provided bounds for $E_{\mathrm{paths}}$ assuming $p(x_i,x_{i+1})$ are independent. In our model, $p(x_i,x_{i+1})$ are not strictly independent. However, when the size of a region is sufficiently large, the distances $\|x_1-x_2\|, \|x_2-x_3\|, \cdots, \|x_{j-1}-x_j\|$ become approximately independent and identically distributed, in which case we can approximate 
\begin{align}
    &\mathbb{E}_{x_1,x_2,\cdots x_j}[p(x_1,x_2)p(x_2,x_3)\cdots p(x_{j-1},x_j)] \notag\\
    &\approx \mathbb{E}_{x_1,x_2}[p(x_1,x_2)]^{j-1}.
\end{align}
To verify the approximate independence, we compute the correlation coefficient via Monte Carlo simulation with 9 million samples, which shows that in a Waxman network with $R=100~km$ and $m=1$,
\begin{align}
    \rho(p(x_1,x_2), p(x_2,x_3)) \approx 0.0337 \pm 0.001.
\end{align}
Let $E\coloneqq\mathbb{E}_{x_1,x_2}[p(x_1,x_2)]$. Using the results in Ref. \cite{van2001paths}, we conclude that
\begin{align}
    E_\mathrm{paths} \approx \tilde{E}_\mathrm{paths} \coloneqq \sum_{j=1}^{N-1}\frac{(N-2)!}{(N-j-1)!} E^j,
\end{align}
which can be bounded as follows \cite{van2001paths}:
\begin{align}
    &(N-2)!E^{N-1}e^{1/E}-\frac{EN}{(EN-1)(N-1)} \notag\\
    &\quad< \tilde{E}_\mathrm{paths} \notag\\ 
    &\quad< (N-2)!E^{N-1}e^{1/E} -\frac{EN+1}{EN(N-1)}.
\end{align}
Note that for large $N$, the lower and upper bounds coincide up to $O(1/N)$:
\begin{align}
    \tilde{E}_\mathrm{paths} = (N-2)!E^{N-1}e^{1/E} + O(1/N).
\end{align}
Additionally, when
\begin{align}
    E \gtrsim \frac{1}{N}\left(1+\sqrt{\frac{2}{N}\ln\frac{N}{2\pi}}\left[1+O(1/N^\sigma)\right]\right),
\end{align}
for some undetermined $\sigma\in(0,1)$, the expected number of paths $\tilde{E}_\mathrm{paths}\geq1$; that is, the expected number of paths between any two nodes is at least 1. This signals that the network becomes sufficiently connected.

\subsection{Scale-Free Networks}

The results for scale-free networks are shown in Fig.~\ref{fig:sf-results}. The routing strategies do not show clear distance dependence. In terms of the average rate $\langle R\rangle$, if we assume GHZ measurements have uniform success probability, then GHZ routing achieves higher rate than BSM routing does. This advantage is lost if we make more realistic assumptions, as exponential-decay GHZ routing and (2,3)-GHZ routing performs significantly worse than BSM routing. Instead, under realistic assumptions, it is best to perform the hybrid GHZ-BSM protocol, which achieves rates comparable to BSM routing does when the number of nodes in the network is sufficiently high.

\subsection{SURFnet}

The results for SURFnet are shown in Fig.~\ref{fig:surfnet-results}. The GHZ-based strategies show some evidence of distance independence for large $q$ (Fig. \ref{fig:surfnet-rate-q-0.9}). For $q=0.7$ and $q=0.8$, the rates for all strategies including BSM routing decreases significantly as distance increases. The average rates $\langle R\rangle$ achieved by uniform-success GHZ routing and BSM routing are comparable, while hybrid GHZ-BSM routing performs slightly worse. Curiously, (2,3)-GHZ routing achieves almost 0 rate.

\section{Discussion}
We analyze the performance of GHZ routing strategies in complex networks. Previous literature assumed that $k$-qubit GHZ measurements succeed with a uniform probability $q$ for all $k$. We relax this assumption by considering (i) a model in which the success probability decreases exponentially with $k$ as $q^{k-1}$, and (ii) a setting restricted to two- and three-qubit measurements. We also introduce a hybrid GHZ-BSM routing strategy. Through numerical simulations, we find that GHZ-based routing strategies can achieve end-to-end rates that are largely distance-independent in square grid, Waxman, and SURFnet networks under suitable conditions. In terms of average rate, GHZ routing outperforms conventional BSM routing in square grid and scale-free networks, performs comparably in SURFnet, and performs significantly worse in Waxman networks under the assumption of uniform GHZ measurement success probability. When this assumption is relaxed, GHZ routing becomes inefficient, and the hybrid GHZ-BSM strategy is preferable; however, it only outperforms BSM routing in square grid networks. It is worth emphasizing that GHZ routing does not rely on global information. As discussed in Section~\ref{sect:waxman-phase-transition}, simple network segmentation can significantly improve the average rate of GHZ routing in Waxman networks, increasing it from $O(1)$ to $O(n)$. We expect that similar ideas can be applied to other network classes, further enhancing achievable rates through more sophisticated design.

Much follow-up work remains. From a numerical simulation standpoint, it will be interesting to analyze the performance of these protocols on other physical network topologies or use discrete-event simulators for more realistic implementations. From a more theoretical perspective, it will be useful to establish information-theoretic lower and upper bounds on the achievable rates.

\section*{Acknowledgments}
This material is based upon work supported by the U.S.~Department of Energy Office of Science National Quantum Information Science Research Centers as part of the Q-NEXT Center.
This work was also supported by the Advanced Scientific Computing Research program under contract number DE-AC02-06CH11357 as part of the InterQnet quantum networking project. 
X.C.~and J.L.~were supported through Q-NEXT, while C.Z.~and J.C.~were supported through the InterQnet project. X.C. acknowledges Xiaojuan Ma and Eric Chitambar for the valuable initial discussions.

\bibliographystyle{IEEEtran}
\bibliography{refs}

@article{patil2022entanglement,
  title={Entanglement generation in a quantum network at distance-independent rate},
  author={Patil, Ashlesha and Pant, Mihir and Englund, Dirk and Towsley, Don and Guha, Saikat},
  journal={npj Quantum Information},
  volume={8},
  number={1},
  pages={51},
  year={2022},
  publisher={Nature Publishing Group UK London},
  doi = {10.1038/s41534-022-00536-0},
  url = {https://doi.org/10.1038/s41534-022-00536-0},
}

@article{yook2002modeling,
  title={Modeling the {Internet's} large-scale topology},
  author={Yook, Soon-Hyung and Jeong, Hawoong and Barab{\'a}si, Albert-L{\'a}szl{\'o}},
  journal={Proceedings of the National Academy of Sciences},
  volume={99},
  number={21},
  pages={13382--13386},
  year={2002},
  publisher={National Academy of Sciences},
  doi={10.1073/pnas.172501399},
  url={https://doi.org/10.1073/pnas.172501399}
}

@article{waxman1988routing,
  title={Routing of multipoint connections},
  author={Waxman, Bernard M},
  journal={IEEE Journal on Selected Areas in Communications},
  volume={6},
  number={9},
  pages={1617--1622},
  year={1988},
  publisher={IEEE},
  doi={10.1109/49.12889},
  url={https://doi.org/10.1109/49.12889}
}

@article{van2001paths,
doi = {10.1017/S0269964801154070},
url = {https://doi.org/10.1017/S0269964801154070},
  title={Paths in the simple random graph and the {Waxman} graph},
  author={Van Mieghem, Piet},
  journal={Probability in the Engineering and Informational Sciences},
  volume={15},
  number={4},
  pages={535--555},
  year={2001},
  publisher={Cambridge University Press}
}

@ARTICLE{lakhina2003geographic,
  author={Lakhina, A. and Byers, J.W. and Crovella, M. and Matta, I.},
  journal={IEEE Journal on Selected Areas in Communications}, 
  title={On the geographic location of Internet resources}, 
  year={2003},
  volume={21},
  number={6},
  pages={934-948},
  url = {https://doi.org/10.1109/JSAC.2003.814667},
  doi={10.1109/JSAC.2003.814667}}

@inproceedings{durairajan2015intertubes,
  title={{InterTubes: A study of the US long-haul fiber-optic infrastructure}},
  author={Durairajan, Ramakrishnan and Barford, Paul and Sommers, Joel and Willinger, Walter},
  booktitle={Proceedings of the 2015 ACM Conference on Special Interest Group on Data Communication},
  pages={565--578},
  year={2015},
  url = {https://dx.doi.org/10.1145/2785956.2787499},
  doi = {10.1145/2785956.2787499}
}

@article{harney2025practical,
 title={Practical routing and criticality in large-scale quantum communication networks},
 volume={7},
 ISSN={2643-1564},
 url={http://dx.doi.org/10.1103/vy37-28jc},
 DOI={10.1103/vy37-28jc},
 number={4},
 journal={Physical Review Research},
 publisher={American Physical Society (APS)},
 author={Harney,
 Cillian and Pirandola,
 Stefano},
 year={2025},
 month=nov 
}

@article{barabasi1999emergence,
  title={Emergence of scaling in random networks},
  author={Barab{\'a}si, Albert-L{\'a}szl{\'o} and Albert, R{\'e}ka},
  journal={Science},
  volume={286},
  number={5439},
  pages={509--512},
  year={1999},
  publisher={American Association for the Advancement of Science},
  doi={10.1126/science.286.5439.509},
  url={https://doi.org/10.1126/science.286.5439.509}
}

@INPROCEEDINGS{kaur2023distribution,
  author={Kaur, Eneet and Guha, Saikat},
  booktitle={2023 IEEE International Conference on Quantum Computing and Engineering (QCE)}, 
  title={Distribution of entanglement in two-dimensional square grid network}, 
  year={2023},
  volume={01},
  number={},
  pages={1154--1164},
  url = {https://doi.org/10.1109/QCE57702.2023.00130},
  doi={10.1109/QCE57702.2023.00130}}

@article{pant2019routing,
  title={Routing entanglement in the quantum internet},
  author={Pant, Mihir and Krovi, Hari and Towsley, Don and Tassiulas, Leandros and Jiang, Liang and Basu, Prithwish and Englund, Dirk and Guha, Saikat},
  journal={npj Quantum Information},
  volume={5},
  number={1},
  pages={25},
  year={2019},
  publisher={Nature Publishing Group UK London},
  doi={10.1038/s41534-019-0139-x},
  url={https://doi.org/10.1038/s41534-019-0139-x}
}

@article{ekert1991quantum,
  title={Quantum cryptography based on {Bell’s} theorem},
  author={Ekert, Artur K},
  journal={Physical Review Letters},
  volume={67},
  number={6},
  pages={661},
  year={1991},
  publisher={APS},
  doi={10.1103/PhysRevLett.67.661},
  url={https://doi.org/10.1103/PhysRevLett.67.661}
}

@article{caleffi2024distributed,
  title={Distributed quantum computing: a survey},
  author={Caleffi, Marcello and Amoretti, Michele and Ferrari, Davide and Illiano, Jessica and Manzalini, Antonio and Cacciapuoti, Angela Sara},
  journal={Computer Networks},
  volume={254},
  pages={110672},
  year={2024},
  publisher={Elsevier},
  url = {https://doi.org/10.1016/j.comnet.2024.110672},
  doi = {10.1016/j.comnet.2024.110672},
}

@article{zhang2021distributed,
  doi={10.1088/2058-9565/abd4c3},
  url={https://doi.org/10.1088/2058-9565/abd4c3},
  title={Distributed quantum sensing},
  author={Zhang, Zheshen and Zhuang, Quntao},
  journal={Quantum Science and Technology},
  volume={6},
  number={4},
  pages={043001},
  year={2021},
  publisher={IOP Publishing}
}

@article{abane2025entanglement,
  title={Entanglement routing in quantum networks: A comprehensive survey},
  author={Abane, Amar and Cubeddu, Michael and Mai, Van Sy and Battou, Abdella},
  journal={IEEE Transactions on Quantum Engineering},
  year={2025},
  publisher={IEEE},
  doi = {10.1109/TQE.2025.3541123},
  url = {https://doi.org/10.1109/TQE.2025.3541123},
}

@article{azuma2023quantum,
  title={Quantum repeaters: From quantum networks to the quantum internet},
  author={Azuma, Koji and Economou, Sophia E and Elkouss, David and Hilaire, Paul and Jiang, Liang and Lo, Hoi-Kwong and Tzitrin, Ilan},
  journal={Reviews of Modern Physics},
  volume={95},
  number={4},
  pages={045006},
  year={2023},
  publisher={APS},
  doi={10.1103/RevModPhys.95.045006},
  url={https://doi.org/10.1103/RevModPhys.95.045006}
}

@article{pirandola2019end,
  title={End-to-end capacities of a quantum communication network},
  author={Pirandola, Stefano},
  journal={Communications Physics},
  volume={2},
  number={1},
  pages={51},
  year={2019},
  publisher={Nature Publishing Group UK London},
  doi={10.1038/s42005-019-0147-3},
  url={https://doi.org/10.1038/s42005-019-0147-3}
}

@article{zhuang2021quantum,
  title={Quantum communication capacity transition of complex quantum networks},
  author={Zhuang, Quntao and Zhang, Bingzhi},
  journal={Physical Review A},
  volume={104},
  number={2},
  pages={022608},
  year={2021},
  publisher={APS},
  doi={10.1103/PhysRevA.104.022608},
  url={https://doi.org/10.1103/PhysRevA.104.022608}
}

@article{brito2020statistical,
  title={Statistical properties of the quantum internet},
  author={Brito, Samura{\'\i} and Canabarro, Askery and Chaves, Rafael and Cavalcanti, Daniel},
  journal={Physical Review Letters},
  volume={124},
  number={21},
  pages={210501},
  year={2020},
  publisher={APS},
  doi={10.1103/PhysRevLett.124.210501},
  url={https://doi.org/10.1103/PhysRevLett.124.210501}
}

@article{acin2007entanglement,
  title={Entanglement percolation in quantum networks},
  author={Ac{\'\i}n, Antonio and Cirac, J Ignacio and Lewenstein, Maciej},
  journal={Nature Physics},
  volume={3},
  number={4},
  pages={256--259},
  year={2007},
  publisher={Nature Publishing Group UK London},
url = {https://doi.org/10.1038/nphys549},
doi = {10.1038/nphys549},
}

@misc{topology-zoo,
    title={The {I}nternet {T}opology {Z}oo},
    oldurl={https://topology-zoo.org/},
    url = {https://github.com/mroughan/InternetTopologyZoo}
    }

@Inbook{grimmett1999percolation,
author="Grimmett, Geoffrey",
title="What is Percolation?",
bookTitle="Percolation",
year="1999",
publisher="Springer Berlin Heidelberg",
address="Berlin, Heidelberg",
pages="1--31",
isbn="978-3-662-03981-6",
doi="10.1007/978-3-662-03981-6_1",
url="https://doi.org/10.1007/978-3-662-03981-6_1"
}

@article{smith2007average,
  doi = {10.48550/arXiv.0710.2947},
  url = {https://doi.org/10.48550/arXiv.0710.2947},
  title={Average path length in complex networks: Patterns and predictions},
  author={Smith, Reginald D},
  journal={arXiv preprint arXiv:0710.2947},
  year={2007}
}

@inproceedings{roughan2019estimating,
url = {https://doi.org/10.1007/978-3-030-25070-6_6},
doi = {10.1007/978-3-030-25070-6_6},
  title={Estimating the parameters of the {Waxman} random graph},
  author={Roughan, Matthew and Tuke, Jonathan and Parsonage, Eric},
  booktitle={International Workshop on Algorithms and Models for the Web-Graph},
  pages={71--86},
  year={2019},
  organization={Springer}
}

@article{bartolucci2023fusion,
url = {https://doi.org/10.1038/s41467-023-36493-1},
doi = {10.1038/s41467-023-36493-1},
  title={Fusion-based quantum computation},
  author={Bartolucci, Sara and Birchall, Patrick and Bombin, Hector and Cable, Hugo and Dawson, Chris and Gimeno-Segovia, Mercedes and Johnston, Eric and Kieling, Konrad and Nickerson, Naomi and Pant, Mihir and others},
  journal={Nature Communications},
  volume={14},
  number={1},
  pages={912},
  year={2023},
  publisher={Nature Publishing Group UK London}
}

@article{calsamiglia2001maximum,
  title={Maximum efficiency of a linear-optical {Bell}-state analyzer},
  author={Calsamiglia, John and L{\"u}tkenhaus, Norbert},
  journal={Applied Physics B},
  volume={72},
  number={1},
  pages={67--71},
  year={2001},
  publisher={Springer},
  url = {https://doi.org/10.1007/s003400000484},
  doi = {10.1007/s003400000484}
}

@article{grice2011arbitrarily,
  title={Arbitrarily complete {Bell}-state measurement using only linear optical elements},
  author={Grice, Warren P},
  journal={Physical Review A—Atomic, Molecular, and Optical Physics},
  volume={84},
  number={4},
  pages={042331},
  year={2011},
  publisher={APS},
  url = {https://doi.org/10.1103/PhysRevA.84.042331},
  doi = {10.1103/PhysRevA.84.042331}
}

\end{document}